\documentclass[runningheads]{llncs}
\usepackage[T1]{fontenc}
\usepackage{graphicx}
\usepackage[colorlinks=true, linkcolor=blue, citecolor=blue, urlcolor=blue]{hyperref}
\usepackage{orcidlink}
\begin{document}
\titlerunning{Building Regulation Capacity in Human--AI Collaborative Learning}
\title{Building Regulation Capacity in Human–AI Collaborative Learning: A Human-Centred GenAI System}
%
%
\author{
Yujing Zhang\inst{1}\orcidlink{0009-0009-3324-2501} \and
Jionghao Lin\inst{1}\thanks{Corresponding author.}\orcidlink{0000-0003-3320-3907}
}

\institute{
The University of Hong Kong, Hong Kong, China \\
\email{zhangyujing@connect.hku.hk}
\and
Carnegie Mellon University, Pittsburgh, PA, USA
\and
Monash University, Clayton, VIC, Australia
}

\institute{
The University of Hong Kong, Hong Kong SAR, China \\
\email{zhangyujing@connect.hku.hk}
}

\authorrunning{Y. Zhang and J. Lin}

\maketitle              

\begin{abstract}

Collaborative learning works when groups regulate together by setting shared goals, coordinating participation, monitoring progress, and responding to breakdowns through co-regulation (CoRL) and socially shared regulation (SSRL). As generative AI (GenAI) enters group work, however, it remains unclear whether and how it supports these socially distributed regulation processes. This doctoral project proposes a GenAI-supported collaborative learning system grounded in CoRL and SSRL to strengthen groups' \emph{socially distributed regulation capacity}. The system links three components: (1) group activity generation; (2) an in-group support agent that provides process-focused prompts without giving solutions; (3) and an embedded learning analytics dashboard that turns interaction traces into timely summaries for monitoring and decision making. The project progresses from mechanism to design to impact: it first identifies how GenAI reshapes regulation patterns and which patterns indicate more effective Human–AI collaboration, then builds an integrated GenAI system that targets these patterns, and finally evaluates whether the GenAI system improves regulation capacity and group performance across varying levels of GenAI involvement. Expected contributions include a teacher-in-the-loop system for Human–AI collaboration and process-level evidence on how GenAI reconfigures CoRL and SSRL in group work.

\end{abstract}
\section{Introduction}

\begin{figure}[t]
    \centering
    \includegraphics[width=\linewidth]{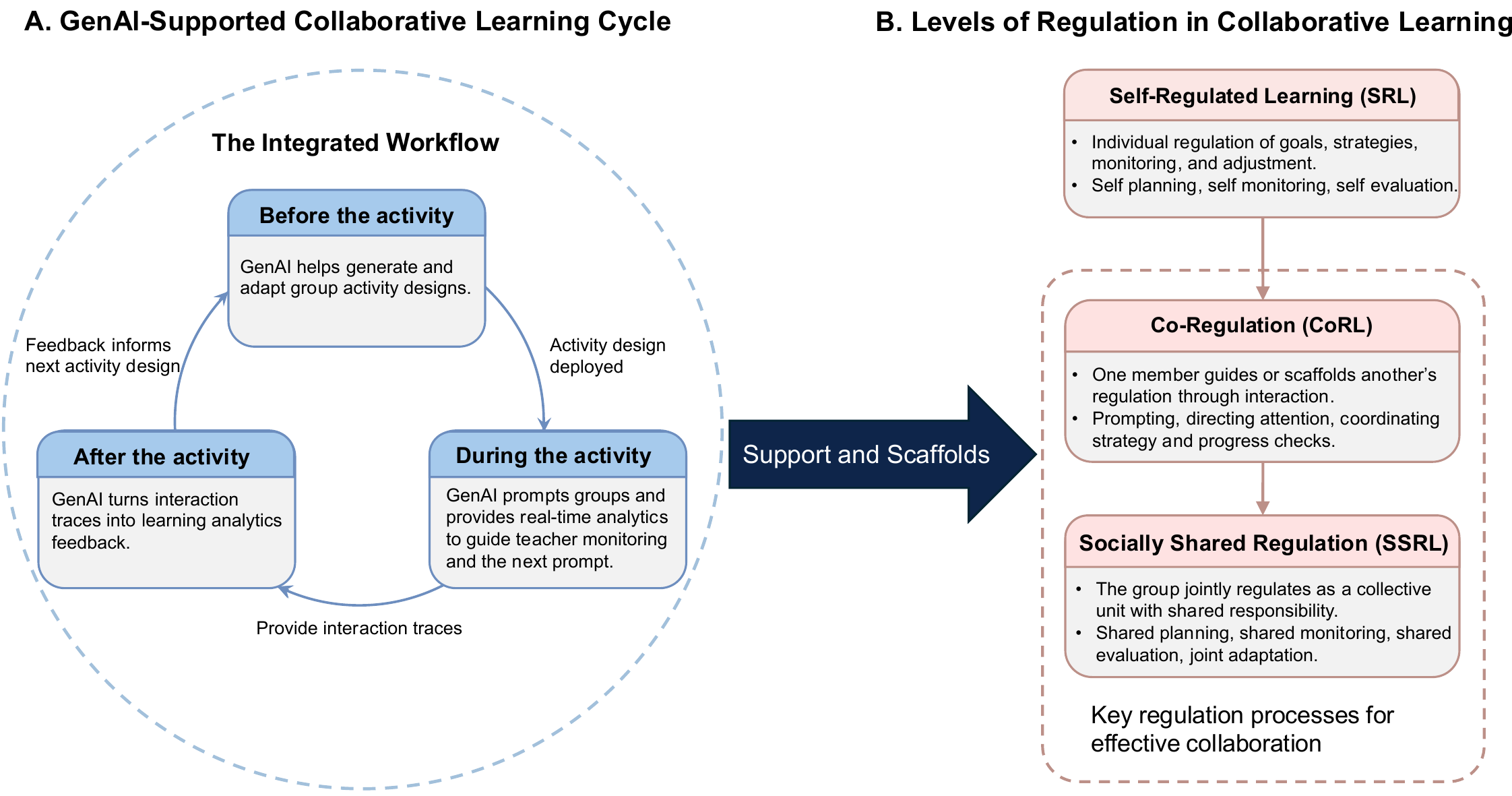} 
    \caption{Focus of the doctoral study: (A) a GenAI-supported CSCL cycle designed to strengthen socially distributed regulation, grounded in (B) SRL, CoRL, and SSRL theory.}
    \label{fig:workflow}
\end{figure}

As collaboration becomes a critical 21st-century competency, Computer-Supported Collaborative Learning (CSCL) has emerged as a key pedagogical approach that leverages digital technologies to support group reasoning and shared knowledge construction \cite{baker-2024}. In parallel, the rise of generative AI (GenAI) introduces new design possibilities for CSCL. GenAI refers to AI models that generate content (e.g., text, images, or video) from natural-language prompts and can engage in natural back-and-forth conversation by interpreting context and producing coherent responses \cite{Cao2023}. It can also follow structured instructions and summarise complex information into concise, usable outputs \cite{Cao2023}. Thus, these capabilities make GenAI well suited to support an integrated CSCL cycle. As illustrated in Fig.~\ref{fig:workflow}A, GenAI can, in principle, support an integrated workflow across three phases of a CSCL activity that closes the loop from group activity design to in-group support and learning analytics, and back to evidence-informed redesign. In the \textit{Before the activity} phase, GenAI can support teachers in generating and adapting a structured group activity design based on their goals and constraints \cite{karaman2024lesson}. This design is then deployed to the group activity. In the \textit{During the activity} phase, GenAI supports collaboration in two coordinated ways: it can prompt groups with process-focused messages that encourage peer sensemaking \cite{yan2024genai}; and it can support real-time learning analytics by interpreting unstructured data such as collaborative dialogues and text records to produce simple indicators of group progress, engagement, and performance \cite{ouyang2024ai}. These indicators guide both teacher attention and the timing and type of the next prompt. The activity produces interaction traces, which are captured for analysis. In the \textit{After the activity} phase, GenAI summarises these traces into teacher-facing learning analytics feedback, which directly informs revisions to the next activity design \cite{ouyang2024ai,yan2024genai}. However, existing research and tools often treat activity generation, in-activity support, and learning analytics as separate functions, and integrated systems that connect them into a single loop remain rare. Consequently, teachers face high coordination costs to use these tools together, which can prevent them from iteratively improving and reusing collaborative activities in authentic classroom contexts.

Additionally, to unlock the full potential of our proposed integrated workflow (Fig.~\ref{fig:workflow}B), it is necessary to strengthen the processes that make collaboration effective. In CSCL, productive collaboration is defined by more than participation alone; it depends on whether groups can sustain joint work over time through shared planning, monitoring, evaluation, adaptation, and repair when obstacles arise \cite{jarvela2016socially}. To explain these processes, current research extends self-regulated learning (SRL) beyond the individual to socially distributed forms of regulation, including co-regulation (CoRL), where learners guide one another’s regulation through interaction, and socially shared regulation (SSRL), where regulation is enacted collectively at the group level \cite{hadwin2018regulation} (Fig.\ref{fig:workflow}B). As such, these socially distributed forms of regulation are central to effective collaboration because they emerge through coordination, mutual orientation, and shared responsibility among group members. Nevertheless, much GenAI-mediated collaboration research has paid limited attention to these regulation capacities, leaving open the question of how GenAI should be designed to strengthen, rather than displace, shared regulation in group work.

To address these gaps, this doctoral research aims to build an integrated GenAI-supported CSCL system to strengthen socially distributed regulation (CoRL and SSRL). The research is organised around three \textbf{Research Questions} that progress from mechanism to design to evaluation:
\begin{itemize}
    \item \textbf{RQ1:} How does GenAI reconfigure socially distributed regulation (CoRL and SSRL) and group interaction patterns in group work, and which patterns distinguish more effective Human--AI collaboration?
    \item \textbf{RQ2:} How can we design and implement an integrated GenAI-supported CSCL system that links group activity generation, in-group support during the activity, and a real-time learning analytics dashboard?
    \item \textbf{RQ3:} To what extent can this GenAI-supported system strengthen socially distributed regulation (CoRL and SSRL) and improve group performance?
\end{itemize}

\begin{figure}[h!]
    \centering
    \includegraphics[width=\linewidth]{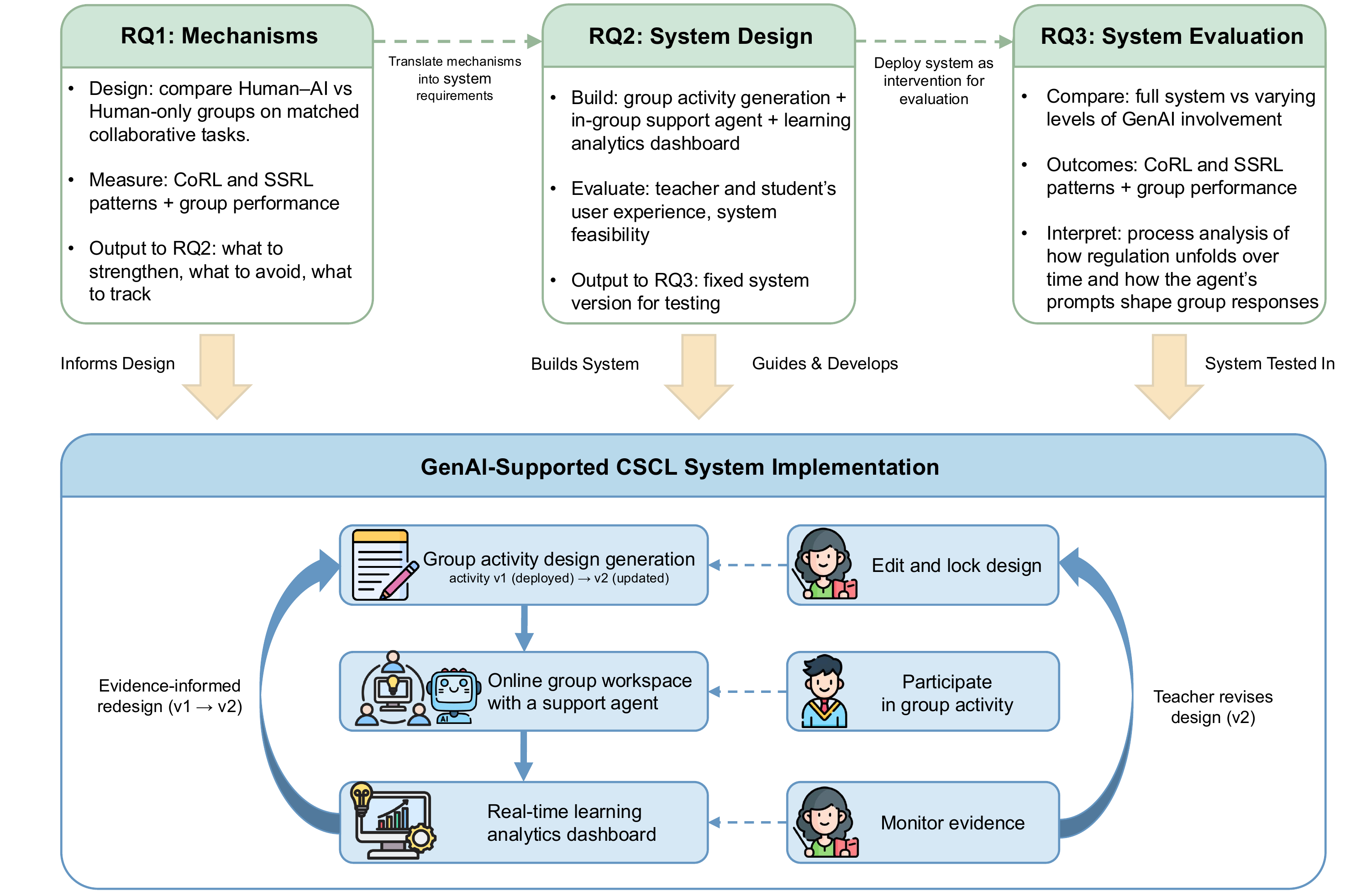}
    \caption{Overview linking RQ1–RQ3 to the GenAI collaboration system.}
    \label{fig:method_overview}
\end{figure}

Fig.~\ref{fig:method_overview} summarises the \textbf{RQ1}--\textbf{RQ3}  progression (top) and the proposed teacher-in-the-loop system linking activity generation, in-group support, and real-time learning analytics in an evidence-informed loop (bottom).

\section{Background and Theoretical Basis}
\label{sec:background}

This study is grounded in regulation learning theory, specifically \emph{co-regulation} (CoRL) and \emph{socially shared regulation} (SSRL) (See Fig.\ref{fig:workflow}B), to guide the design of the proposed GenAI-supported system. CoRL refers to episodes where one learner temporarily scaffolds another learner’s regulation, whereas SSRL refers to moments when regulation is enacted collectively at the group level \cite{hadwin2018regulation}. Together, these constructs capture how groups regulate through interaction by organising goals and strategies and by monitoring, evaluating, adapting, and repairing their joint work over time \cite{hadwin2018regulation,rogat2011ssrl}. Therefore, this project treats GenAI system as \emph{bounded support}, constrained to regulation-oriented moves intended to strengthen CoRL and SSRL. Accordingly, CoRL and SSRL serve both as evaluation targets and as design specifications for what the system prompts, when it intervenes, and which trace-based indicators it surfaces to teachers for monitoring and improvement \cite{edwards2025human}.

\section{Methodology and Research Plan}

\subsection{Overview}
Fig.\ref{fig:method_overview} summarises the staged plan linking \textbf{RQ1}–\textbf{RQ3} to system implementation. \textbf{RQ1} identifies regulation mechanisms and yields design targets and indicators. \textbf{RQ2} builds and pilots the integrated system, producing a fixed configuration for evaluation. \textbf{RQ3} evaluates system impact.

\subsection{Current Progress: RQ1 Regulation mechanisms in Human--AI group work}
To date, we have completed the empirical work for RQ1. We conducted a parallel-group randomised experiment with university students ($N=71$), working in triads, comparing Human-only and Human--AI groups on matched collaborative tasks. In the Human-only condition, groups used Microsoft Teams with no GenAI. In the Human–AI condition, groups used an online CSCL platform with an embedded GenAI agent (\texttt{OpenAI GPT-4o mini}) for real-time text collaboration. The agent did not give content or solutions; it delivered process-focused prompts to elicit explanation, encourage perspective-taking, and support planning and monitoring. We used statistical comparisons and network analysis to examine group-level regulation and how regulatory processes were enacted through participation moves across conditions.

\subsubsection{Key findings}
GenAI availability reconfigured collaborative regulation. Compared with Human-only groups, Human--AI groups shifted from predominantly socially shared regulation towards more hybrid co-regulatory forms, with selective increases in directive, obstacle-oriented, and affective processes. While overall participation-focus distributions were broadly similar across conditions, Human--AI groups exhibited a more differentiated regulatory structure, with regulation more tightly coupled to monitoring, evaluative reasoning, and reporting. By contrast, Human-only groups centred regulation on shared strategic negotiation and shared action change, supported by a wider range of participatory actions. Group performance did not differ by condition; however, only high-performing Human--AI teams showed a coherent obstacle-driven plan--monitor pattern.

\subsubsection{Implications.}
Together, these findings indicate that GenAI can shift where regulatory responsibility sits and how regulation is coordinated in group work. Consequently, effective Human–AI collaboration is unlikely to emerge by default. GenAI should therefore scaffold shared regulation by prompting joint planning, progress checks, and coordinated action when obstacles arise. During group work, learning analytics can use indicators such as missing or weak coupling patterns to trigger targeted support and reorient groups towards effective regulation. 

\subsection{Future Work}

\noindent \textbf{Human-centred system design and feasibility (\textbf{RQ2}).} Building on the \textbf{RQ1} evidence, \textbf{RQ2} translates the identified targets into system requirements for activity structure, bounded in-group prompting, and trace-based indicators surfaced to teachers. We will answer \textbf{RQ2} by building GenAI-supported CSCL system and piloting it with teachers and students to assess usability and feasibility.  This system has three components: group activity generation, an in-group support agent during the activity, and a real-time learning analytics dashboard (See Figure~\ref{fig:method_overview}). We will assess feasibility and perceived value through teacher and student feedback, including semi-structured interviews, alongside interface logs and system metrics (e.g., bound compliance and dashboard latency). These findings will guide system refinement and finalise the indicators used in RQ3.

\noindent \textbf{System impact under different levels of GenAI involvement (\textbf{RQ3}).} We will answer \textbf{RQ3} by conducting a quasi-experimental, between-groups comparison with matched tasks and settings, varying the level of GenAI involvement. Study~3 will compare three conditions: (i) \textit{baseline}, with unstructured chatbot-only access during group work; (ii) \textit{in-group support only}, where the GenAI agent provides bounded, process-focused prompts during the activity; and (iii) \textit{in-group support plus teacher-monitoring learning analytics dashboard}, where the same in-group support is paired with a dashboard for teacher monitoring and decision making. Outcomes will include trace-based indicators of collaborative regulation capacity and group performance, supplemented by process analysis of group responses to prompts and regulation dynamics over time to interpret differences across conditions.

\section{Expected Contributions}

This project will contribute to Artificial Intelligence in Education (AIED) in three ways. \textit{First}, it will deliver a closed-loop GenAI-supported CSCL system aligned with the collaborative learning cycle (see Fig.~\ref{fig:workflow}A), linking teacher-facing group activity generation, an in-group support agent during the activity, and trace-based learning analytics feedback for monitoring and improvement.

\textit{Second}, it will contribute human-centred design knowledge by placing teachers in the system loop (See Fig.~\ref{fig:method_overview}). Teachers remain accountable through key control points: edit and lock the activity design before class, monitor dashboard evidence during the activity, and revise the design based on trace-based feedback after the activity, clarifying how control points and evidence presentation support appropriate trust.

\textit{Third}, it will advance AIED theory of Human–AI collaboration by providing process-level evidence on how GenAI reconfigures co-regulation and socially shared regulation, and when these shifts support productive collaboration.

\bibliographystyle{splncs04}
\bibliography{references}

@article{baker-2024,
	author = {Baker, Michael and Reimann, Peter},
	journal = {International Journal of Computer-Supported Collaborative Learning},
	month = {8},
	number = {3},
	pages = {273--281},
	title = {{CSCL: a learning and collaboration science?}},
	volume = {19},
	year = {2024},
}

@article{ouyang2024ai,
  title={AI-driven learning analytics applications and tools in computer-supported collaborative learning: A systematic review},
  author={Ouyang, Fan and Zhang, Liyin},
  journal={Educational Research Review},
  volume={44},
  pages={100616},
  year={2024},
  publisher={Elsevier}
}

@article{karaman2024lesson,
  title={Are Lesson Plans Created by ChatGPT More Effective? An Experimental Study.},
  author={Karaman, Muhammet Remzi and others},
  journal={International Journal of Technology in Education},
  volume={7},
  number={1},
  pages={107--127},
  year={2024},
  publisher={ERIC}
}

@incollection{hadwin2018regulation,
  author    = {Hadwin, Allyson and J{\"a}rvel{\"a}, Sanna and Miller, Mariel},
  title     = {Self-regulation, co-regulation, and shared regulation in collaborative learning environments},
  booktitle = {Handbook of Self-Regulation of Learning and Performance},
  editor    = {Schunk, Dale H. and Greene, Jeffrey A.},
  edition   = {2},
  pages     = {83--106},
  publisher = {Routledge/Taylor \& Francis Group},
  year      = {2018},
}

@article{jarvela2016socially,
  title={Socially shared regulation of learning in CSCL: Understanding and prompting individual-and group-level shared regulatory activities},
  author={J{\"a}rvel{\"a}, Sanna and Kirschner, Paul A and Hadwin, Allyson and J{\"a}rvenoja, Hanna and Malmberg, Jonna and Miller, Mariel and Laru, Jari},
  journal={International Journal of Computer-Supported Collaborative Learning},
  volume={11},
  number={3},
  pages={263--280},
  year={2016},
  publisher={Springer}
}

@article{rogat2011ssrl,
  author  = {Rogat, Tobin K. and Linnenbrink-Garcia, Lisa},
  title   = {Socially Shared Regulation in Collaborative Groups: An Analysis of the Interplay Between Quality of Social Regulation and Group Processes},
  journal = {Cognition and Instruction},
  year    = {2011},
  volume  = {29},
  number  = {4},
  pages   = {375--415},
}

@article{yan2024genai,
  author  = {Yan, L. and Greiff, S. and Teuber, Z. and Ga{\v{s}}evi{\'c}, D.},
  title   = {Promises and challenges of generative artificial intelligence for human learning},
  journal = {Nature Human Behaviour},
  year    = {2024},
  volume  = {8},
  pages   = {1839--1850}
}

@misc{Cao2023,
  author       = {Cao, Y. and Li, S. and Liu, Y. and Yan, Z. and Dai, Y. and Yu, Philip S. and Sun, L.},
  title        = {A comprehensive survey of AI-generated content (AIGC): A history of generative AI from GAN to ChatGPT},
  year         = {2023},
  howpublished = {arXiv preprint},
  eprint       = {2303.04226}
}

@article{edwards2025human,
  title={Human-AI collaboration: Designing artificial agents to facilitate socially shared regulation among learners},
  author={Edwards, Justin and Nguyen, Andy and L{\"a}ms{\"a}, Joni and Sobocinski, Marta and Whitehead, Ridwan and Dang, Belle and Roberts, Anni-Sofia and J{\"a}rvel{\"a}, Sanna},
  journal={British Journal of Educational Technology},
  volume={56},
  number={2},
  pages={712--733},
  year={2025},
  publisher={Wiley Online Library}
}

\end{document}